\documentclass[conference]{IEEEtran}
\usepackage{graphics}
\usepackage{graphicx}
\usepackage{epsfig}
\usepackage{amsmath,amssymb}
\usepackage{mathrsfs,amsfonts,fancyhdr}
\usepackage{multirow}
\usepackage{cite}

\begin{document}

\title{A Half-Duplex Cooperative Scheme with Partial Decode-Forward Relaying}

\author{\authorblockN{Ahmad Abu Al Haija, and Mai Vu,\\}
\authorblockA{Department of Electrical and Computer Engineering\\
McGill University\\
Montreal, QC H3A 2A7\\
Emails: ahmad.abualhaija@mail.mcgill.ca, mai.h.vu@mcgill.ca}}

\maketitle

\begin{abstract}
In this paper, we present a new cooperative communication scheme consisting of two users in half-duplex mode communicating with one destination over a discrete memoryless channel. The users encode messages in independent blocks and divide the transmission of each block into 3 time slots with variable durations. Cooperation is performed by partial decode-forward relaying over these 3 time slots. During the first two time slots, each user alternatively transmits and decodes, while during the last time slot, both users cooperate to send information to the destination. An achievable rate region for this scheme is derived using superposition encoding and joint maximum likelihood (ML) decoding across the 3 time slots. An example of the Gaussian channel is treated in detail and its achievable rate region is given explicitly. Results show that the proposed half-duplex scheme achieves significantly larger rate region than the classical multiple access channel and approaches the performance of a full-duplex cooperative scheme as the inter-user channel quality increases.
\end{abstract}


\IEEEpeerreviewmaketitle

\section{Introduction}\label{sec:intro}

\IEEEPARstart{C}{ooperative} communication has received much attention because of its ability to increase throughput and reliability of communication systems. Cooperation even in the simplest scenarios between two users can be modeled in different ways: as a multiple access channel (MAC) in various forms (with common message, conferencing encoders or generalized feedback) or as a relay channel. Slepian-Wolf studied the capacity region for the MAC with common message \cite{haykin2005crb} and Willems for a MAC with conferencing encoders \cite{ghasemi2007fls}. In \cite{fcc2005fof}, Willems et al. derived the achievable rate region for the MAC with generalized feedback using block Markov encoding and backward decoding. In their scheme, each user uses the feedback link to improve their achievable rates. Sendonaris et al. \cite{cognitiveRT}  applied this coding scheme into cellular networks operating over fading channels and showed that cooperation leads to a rate region larger than the classical MAC. The effects of cooperation on the secrecy of the MAC with generalized feedback have been illustrated in \cite{lieblein1955mos}. The relay channel introduced by Van der Meulen in \cite{ray1}, on the other hand, models a different way of cooperation.  Cover and El Gamal \cite{ray2} derived its achievable rate region under different relaying schemes such as decode-forward, partial decode-forward, and compress-forward. Kramer et al.  \cite{kolodzy2006itm} generalized these schemes to a relay network. In all these cooperative coding schemes, the communication is assumed to be full-duplex.

Recently, half-duplex communication has received increasing attention because of its practical application for example in wireless. Existing works include performance analysis of half-duplex cooperative schemes in terms of  outage capacity \cite{sagias2004pad}, outer bounds for the capacity of half-duplex relay \cite{simon2005dco} and interference channels \cite{nakagami1960mdg}. The relay channel with orthogonal transmitting components which models frequency division, has its capacity established in \cite{rorth}.

In this paper, we explicitly take the half-duplex constraint into account in designing codes for cooperative communication. We propose a new cooperative half-duplex scheme that combines ideas of the MAC with generalized feedback and relay channels with partial decode-forward relaying. In the MAC with generalized feedback, both users have their own information to send and they cooperate by retransmitting what they received from each other to the destination. Therefore, the relay channel can be seen as a MAC with generalized feedback when one of the users just relays the other user's information without its own information to send.

In our scheme, each of the two users has
its own information to be transmitted to the destination. They cooperate using partial decode-forward relaying in order to improve their rates. However, unlike \cite{fcc2005fof,cognitiveRT}, each user works in half-duplex mode. In order to ensure this half-duplex constraint, we use time division and divide each of the independent transmission blocks into three time slots with variable durations. While each user alternatively transmits and receives during the first two slots, both of them transmit
during the last one. We employ rate splitting and superposition coding techniques for encoding similar to \cite{fcc2005fof,cognitiveRT}, but the transmission is performed in independent blocks without any block Markovity. As a consequence, instead of backward decoding\cite{fcc2005fof}, decoding is done independently at the end of each block by using joint ML
decoding \cite{abramowitz1972hmf}  over all three time slots. This difference is important for practical systems that have delay constraints.

The remainder of this paper is organized as follows. Section \ref{sec:system_model} describes the channel model. Section \ref{sec:achrr} provides an achievable rate region and explains the coding scheme. The achievable rates for Gaussian channel is provided in Section \ref{sec:cap. gau}. Finally, Section \ref{sec:conclusion} concludes the paper.


\section{Channel Model}\label{sec:system_model}
\begin{figure}[]
    \begin{center}
    \includegraphics[width=0.48\textwidth]{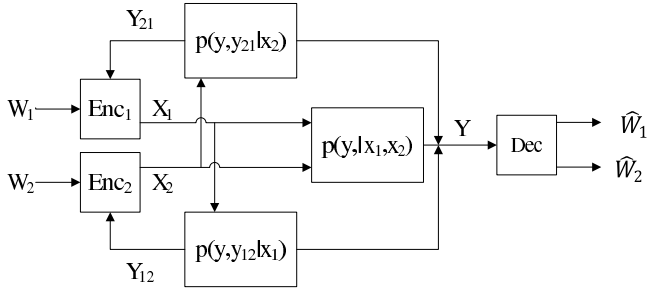}
    \caption{The half-duplex cooperative MAC model.} \label{fig:system_model}
    \end{center}
\vspace*{-7mm}
\end{figure}

The two user discrete memoryless half-duplex cooperative MAC can be defined as follows. Two input alphabets ${\cal X}_1$
and ${\cal X}_2$, three output alphabets ${\cal Y}$, ${\cal Y}_{12}$, and ${\cal Y}_{21}$, and three conditional transition probabilities $p(y|x_1,x_2)$, $p(y,y_{12}|x_1)$, and $p(y,y_{21}|x_2)$ as shown in Fig.\ref{fig:system_model}. This channel is quite similar to the MAC with generalized feedback in \cite{fcc2005fof}. However, an additional requirement is that no two channels occur at the same time in order to satisfy the half-duplex constraint. Because of this requirement, the coding scheme given in \cite{fcc2005fof} can not be applied directly.

A $(\lceil2^{nR_1}\rceil,\lceil2^{nR_2}\rceil,n)$ code for this channel consists of two message sets $W_1=\{1,\ldots,\lceil2^{nR_1}\rceil\}$, and $W_2=\{1,\ldots,\lceil2^{nR_2}\rceil\}$, two encoding functions $f_{1i},f_{2i},\;i=1,\ldots,n$, and one decoding function $g$ defined as
\noindent
\begin{align}
&f_{1i}: W_1\times {\cal Y}_{21}^{i-1}\rightarrow{\cal X}_1,\; {}i=1,\ldots,n \nonumber\\
&f_{2i}: W_2\times {\cal Y}_{12}^{i-1}\rightarrow{\cal X}_2,\; {}i=1,\ldots,n \nonumber\\
&g: {\cal Y}^n \rightarrow W_1\times W_2.
\label{Eq:fR}
\end{align}
\noindent Finally, $P_e$ is the average error probability defined as $P_e=P(g(Y^n)\neq (W_1,W_2))$. A rate pair $(R_1,R_2)$ is said to be achievable if there
exists a $(\lceil2^{nR_1}\rceil,\lceil2^{nR_2}\rceil,n)$ code such that $P_e\rightarrow0$ as $n\rightarrow\infty$. The capacity region is the closure of the set of all achievable rates
$(R_1,R_2)$.
\section{A half-Duplex Coding Scheme}\label{sec:achrr}

A coding scheme for the half-duplex cooperative MAC can be established as follows. The transmission is done in blocks of length $n$. Each block is divided
into three time slots with durations $\alpha_1,$ $\alpha_2$ and $(1-\alpha_1-\alpha_2)$. While the destination is always in receiving mode, each user either transmits or receives during the first two time
slots and both of them transmit during the third slot. We employ rate splitting and superposition coding. Consider the first user; it divides its message, $W_1$, into three parts. The first and the third parts, $W_{10}$ and $W_{13}$, are private and transmitted directly to the destination at rates $R_{10}$ and $R_{13}$, respectively. The second part $W_{12}$ is public and transmitted to
the destination in cooperation with the second user at rate $R_{12}$. The transmission of the second user is similar.
\subsection{Achievable Rate Region}
The achievable rate
region is the convex closure of the rate-tuples $(R_{10},R_{12},R_{13},R_{20},R_{21},R_{23})$ satisfying
\noindent
\begin{align*}
R_{10}\leq&\;\text{min}\left(\alpha_1 I(X_{10};Y_1|U),\alpha_1 I(X_{10};Y_{12}|U)\right)=I_1\\
R_{10}+R_{12}\leq&\; \alpha_1 I(X_{10};Y_{12})=I_2\\
R_{20}\leq&\;\text{min}\left(\alpha_2 I(X_{20};Y_2|V),\alpha_2 I(X_{20};Y_{21}|V)\right)=I_3\\
R_{20}+R_{21}\leq&\; \alpha_2 I(X_{20};Y_{21})=I_4\\
R_{13}\leq&\;\alpha_3 I(X_{13};Y_3|U,V,X_{23})=I_5\\
R_{23}\leq&\;\alpha_3 I(X_{23};Y_3|U,V,X_{13})=I_6\\
R_{13}+R_{23}\leq&\;\alpha_3 I(X_{13},X_{23};Y_3|U,V)=I_7\\
R_{1}+R_{23}\leq&\;\alpha_1 I(X_{10};Y_1)+\alpha_3  I(X_{13},X_{23};Y_3|V)=I_8
\end{align*}
\begin{align}
R_{2}+R_{13}\leq&\;\alpha_2 I(X_{20};Y_2)+\alpha_3  I(X_{13},X_{23};Y_3|U)=I_{9}\nonumber \\
R_1+R_2\leq&\; \alpha_1 I(X_{10};Y_1)+\alpha_2 I(X_{20};Y_2)+ \nonumber \\
&\alpha_3  I(X_{13},X_{23};Y_3)=I_{10}
\label{Eq:rnd}
\end{align}
\noindent for some
\begin{align}\label{ipdis}
P^{\ast}=p(x_{10},u)p(x_{20},v)p(x_{13}|u,v)p(x_{23}|u,v)
\end{align}
where $0\leq\alpha_1+\alpha_2\leq1$ and $\alpha_3=1-\alpha_1-\alpha_2$. Now, by applying Fourier-Motzkin Elimination (FME) to the inequalities in (\ref{Eq:rnd}), the achievable rates in terms of $R_1=R_{10}+R_{12}+R_{13}$ and $R_2=R_{20}+R_{21}+R_{23}$ can be expressed as
\noindent
\begin{align}
R_1&\leq I_2+I_5,\;\;\;\;\;\;\;\;\;\;\;\;\;\;\;\;\;\;\;\;R_2\leq I_4+I_6 \nonumber \\
R_1+R_2&\leq I_7+I_2+I_4,\;\;\;\;R_1+R_2\leq I_{10}\nonumber \\
R_1+R_2&\leq I_4+I_8,\;\;\;\;\;\;\;\;\;\;\;\;\!R_1+R_2\leq I_2+I_9
\label{Eq:rfin}
\end{align}
\subsection{Encoding Technique}
\subsubsection{Codebook generation}
Fix $P^{\ast}$ in (\ref{ipdis}). Generate:
\noindent
\begin{itemize}
\item
$2^{nR_{12}}$ i.i.d sequences $u^{n}(w_{12})\sim\prod_{i=1}^np(u_{i})$
\item
$2^{nR_{21}}$ i.i.d sequences $v^{n}(w_{21})\sim\prod_{i=1}^np(v_{i})$
\end{itemize}
\noindent Then for each $u^{n}(w_{12})$ and each $v^{n}(w_{21})$, generate:
\noindent
\begin{itemize}
\item
$2^{nR_{10}}$ i.i.d sequences $x_{10}^{n}(w_{10},w_{12})\sim\prod_{i=1}^np(x_{10i}|u_i),$
\item
$2^{nR_{20}}$ i.i.d sequences $x_{20}^{n}(w_{20},w_{21})\sim\prod_{i=1}^np(x_{20i}|v_i)$
\end{itemize}
Finally, for each pair $(u^{n}(w_{12}),v^{n}(w_{21}))$, generate:
\noindent
\begin{itemize}
\item
$2^{nR_{13}}$ i.i.d sequences $x_{13}^{n}(w_{13},w_{12},w_{21})\sim\prod_{i=1}^np(x_{13i}|u_i,v_i)$
\item
$2^{nR_{23}}$ i.i.d sequences $x_{23}^{n}(w_{23},w_{12},w_{21})\sim\prod_{i=1}^np(x_{23i}|u_i,v_i)$
\end{itemize}
\begin{figure*}[t]
\normalsize
\begin{center}
\begin{tabular}{|c|c|c|c|}
\hline
{}&$1^\text{st}$ slot with length $\alpha_1 n$ &$2^\text{nd}$ slot with length $\alpha_2 n$&$3^\text{rd}$ slot with length $(1-\alpha_1-\alpha_2)n$\\
\hline
first user Tx&$x_{10}^{\alpha_1 n}(w_{10},w_{12})$&$--$&$x_{13, (\alpha_1+\alpha_2) n +1}^{n}(w_{13},w_{12},\tilde{w}_{21})$\\
\hline
second user Tx&$--$&$x_{20}^{\alpha_2 n}(w_{20},w_{12})$&$x_{23, (\alpha_1+\alpha_2) n +1}^{n}(w_{23},\tilde{w}_{12},w_{21})$\\
\hline
$Y_{21}$&$--$&$(\tilde{w}_{20},\tilde{w}_{21})$&$--$\\
\hline
$Y_{12}$&$(\tilde{w}_{10},\tilde{w}_{12})$&$--$&$--$\\
\hline
\multirow{2}{*}{$Y$}&$Y_1$&$Y_2$&$Y_3$\\
\cline{2-4}
&\multicolumn{3}{c|}{$(\hat{w}_{12},\hat{w}_{21},\hat{w}_{10},\hat{w}_{20},\hat{w}_{13},\hat{w}_{23})$}\\
\hline
\end{tabular}
\\
\vspace*{2mm}
{\small Table I: The encoding and decoding schemes for half duplex cooperative scheme}\\
\vspace*{2mm}
\end{center}
\vspace*{-8mm}
\end{figure*}
\subsubsection{Encoding}
In order to send the message pair $(W_1,W_2)$, the first user sends $x_{10}^{\alpha_1 n}(w_{10},w_{12})$ during the $1^\text{st}$ time slot, while
the second user sends $x_{20}^{\alpha_2 n}(w_{20},w_{12})$ during the $2^\text{nd}$ time slot. At the end of the $1^\text{st}$ and $2^\text{nd}$ time slots, the second user and the first user will have the estimated values $(\tilde{w}_{10},\tilde{w}_{12})$ and $(\tilde{w}_{20},\tilde{w}_{21})$, respectively. Then, the first user
sends $x_{13, (\alpha_1+\alpha_2) n +1}^{n}(w_{13},w_{12},\tilde{w}_{21})$ and the second user sends $x_{23, (\alpha_1+\alpha_2)n +1}^{n}(w_{23},\tilde{w}_{12},w_{21})$ during the last time slot. Hence cooperation occurs via decode-forward relaying. Each user decodes the other user's messages during the first two time slots, then forwards the public part of this message during the third time slot. In addition, each user also sends a private message to the destination.

In our scheme, the private parts of the messages are superimposed on the public parts of the same transmission block. This is different from \cite{fcc2005fof,cognitiveRT} in which they are superimposed on the public parts of the previous block which leads to dependent blocks. Also, we can see that our scheme includes as special cases the classical MAC when $\alpha_1=\alpha_2=0$ and the classical TDMA when $\alpha_1=\alpha_2=0.5$.

\subsection{Decoding Technique}
\subsubsection{Decoding at each user}
At the end of the $1^\text{st}$ ($2^\text{nd}$) time slot, the second first  user uses joint maximum likelihood decoding rule to decode $(w_{10},w_{12}),$ $((w_{20},w_{21}))$. Specifically, for given a received sequence, $y_{21}^{\alpha_2 n}$ (or $y_{12}^{\alpha_1 n}$), the user chooses

\noindent $(\hat{w}_{20},\hat{w}_{21}),$ (or $(\hat{w}_{10},\hat{w}_{12})$) for which:
\noindent
\begin{itemize}
\item
$P(y_{21}^{\alpha_2 n}|x_{20}^{\alpha_2 n}(\hat{w}_{20},\hat{w}_{21}))\geq P(y_{21}^{\alpha_2 n}|x_{20}^{\alpha_2 n}(w_{20},w_{21})),$ for all $(w_{20},w_{21})\neq(\hat{w}_{20},\hat{w}_{21})$
\item
$P(y_{12}^{\alpha_1 n}|x_{10}^{\alpha_1 n}(\hat{w}_{10},\hat{w}_{12}))\geq P(y_{12}^{\alpha_1 n}|x_{10}^{\alpha_1 n}(w_{10},w_{12})),$ for all $(w_{10},w_{12})\neq(\hat{w}_{10},\hat{w}_{12})$.
\end{itemize}
\noindent The users can also decode one part only, but this leads to a smaller rate region for a given input distribution of the Gaussian channel.
\subsubsection{Decoding at the destination}
Since the transmitted blocks are independent, the destination can decode at the end of each block. Using jointly
ML decoding, it decides that message vector $(\hat{w}_{12},\hat{w}_{21},\hat{w}_{10},\hat{w}_{20},\hat{w}_{13},\hat{w}_{23})$ was sent if
\noindent
\begin{align*}
&P(\boldsymbol{y}|\boldsymbol{\hat{x}_1}(\hat{w}_{12},\hat{w}_{10},\hat{w}_{13}),\boldsymbol{\hat{x}_2}(\hat{w}_{12},\hat{w}_{10},\hat{w}_{13}))\geq\\
&P(\boldsymbol{y}|\boldsymbol{x_1}(w_{12},w_{10},w_{13}),\boldsymbol{x_2}(w_{21},w_{20},w_{23}))
\end{align*}
\noindent for all $(w_{12},w_{21},w_{10},w_{20},w_{13},w_{23})$, where $\boldsymbol{y}=(y_1^{\alpha_1 n} y_2^{\alpha_2 n}$ $y_3^{1-\alpha_1-\alpha_2})$ is the received sequence from all three time slots.

Note that joint decoding across all $3$ time slots at the destination is important. If the destination decodes in each time slot separately, the rate region will be smaller. However, the joint decoding complicates error analysis as seen next.

The encoding and decoding at each block can be explained with the help of Table I, where $0\leq\alpha_1+\alpha_2\leq1$. The same achievable rate region can also be obtained using joint typicality decoding instead of ML decoding.

\subsection{Error Analysis}\label{subsec:eranz}
Without loss of generality, assume that the message vector $(w_{12}=w_{21}=w_{10}=w_{20}=w_{13}=w_{23}=1)$ was sent and let $\Im$ be this event.The error events at each user can be analyzed as in \cite{abramowitz1972hmf}. To make these error probabilities approach zero, the rate constraints $(I_2,I_4)$ and the second part of $(I_1,I_3)$ must be satisfied. Now, the error events at the destination are
\noindent
\begin{align*}
P_{E_1}&=P\left[w_{12}=w_{21}=w_{20}=w_{13}=w_{23}=1, w_{10}\neq1|\Im\right]\\
P_{E_2}&=P\left[w_{12}=w_{21}=w_{10}=w_{13}=w_{23}=1, w_{20}\neq1|\Im\right]\\
P_{E_3}&=P\left[w_{12}=w_{21}=w_{13}=w_{23}=1, (w_{10}, w_{20})\neq1|\Im\right]\\
P_{E_4}&=P\left[w_{12}=w_{21}=w_{10}=w_{20}=w_{23}=1, w_{13}\neq1|\Im\right]\\
P_{E_5}&=P\left[w_{12}=w_{21}=w_{20}=w_{23}=1, (w_{13}, w_{10})\neq1|\Im\right]\\
P_{E_6}&=P\left[w_{12}=w_{21}=w_{10}=w_{23}=1, (w_{13}, w_{20})\neq1|\Im\right]\\
P_{E_7}&=P\left[w_{12}=w_{21}=w_{23}=1, (w_{13}, w_{10}, w_{20})\neq1|\Im\right]\\
P_{E_8}&=P\left[w_{12}=w_{21}=w_{10}=w_{20}=w_{13}=1, w_{23}\neq1|\Im\right]\\
P_{E_9}&=P\left[w_{12}=w_{21}=w_{20}=w_{13}=1, (w_{23}, w_{10})\neq1|\Im\right]\\
P_{E_{10}}&=P\left[w_{12}=w_{21}=w_{10}=w_{13}=1, (w_{23}, w_{20})\neq1|\Im\right]\\
P_{E_{11}}&=P\left[w_{12}=w_{21}=w_{13}=1, (w_{23}, w_{10}, w_{20})\neq1|\Im\right]\\
P_{E_{12}}&=P\left[w_{12}=w_{21}=w_{10}=w_{20}=1, (w_{13}, w_{23})\neq1|\Im\right]\\
P_{E_{13}}&=P\left[w_{12}=w_{21}=w_{20}=1, (w_{13}, w_{23}, w_{10})\neq1|\Im\right]\\
P_{E_{14}}&=P\left[w_{12}=w_{21}=w_{10}=1, (w_{13}, w_{23}, w_{20})\neq1|\Im\right]
\end{align*}
\begin{align*}
P_{E_{15}}&=P\left[w_{12}=w_{21}=1, (w_{13}, w_{23}, w_{10}, w_{20})\neq1|\Im\right]\\
P_{E_{16}}&=P\left[w_{21}=w_{20}=1, w_{12}\neq1|\Im\right]\\
P_{E_{17}}&=P\left[w_{21}=1, (w_{12}, w_{20})\neq1|\Im\right]\\
P_{E_{18}}&=P\left[w_{12}=w_{10}=1, w_{21}\neq1|\Im\right]\\
P_{E_{19}}&=P\left[w_{12}=1, (w_{21}, w_{10})\neq1|\Im\right]\\
P_{E_{20}}&=P\left[(w_{12}, w_{21})\neq1|\Im\right]
\end{align*}
\noindent where $(x, y,\ldots)\neq1$ means that $x\neq1, y\neq1, \ldots$.
The upper bounds for these error events are derived in the Appendix and they lead to the rate constraints involving the first part of $(I_1,I_3)$ and $(I_5-I_{10})$ as given in (\ref{Eq:rnd}).
\section{Gaussian Channels}\label{sec:cap. gau}
\subsection{Gaussian Channel Model}
The discrete time channel model for the half-duplex cooperative MAC over AWGN channel can be expressed as
\noindent
\begin{align*}
Y_{12}&=K_{12}X_{10}+Z_1,\;\;\;\;\;\;\;\;\;\;\;\;\;\;\;\;\;\;Y_{21}=K_{21}X_{20}+Z_2\\
Y_{1}&=K_{10}X_{10}+Z_{01},\;\;\;\;\;\;\;\;\;\;\;\;\;\;\;\;\;\;Y_{2}=K_{20}X_{20}+Z_{02}\\
Y_{3}&=K_{10}X_{13}+K_{20}X_{23}+Z_{03},\;&.
\end{align*}
\noindent where $K_{12}$ and $K_{21}$ are the inter-user channel coefficients; $K_{10},$ and $K_{20}$ are the channels coefficients between each user and the destination; $Z_1\sim N(0,N_1), Z_2\sim N(0,N_2),$ and $Z_{0i}\sim N(0,N_0),\; i=1,2,3$. Here $X_{10},$ and $X_{13}$ are the transmitted signals from the first user during the $1^\text{st}$ and $3^\text{rd}$ time slots, respectively. Similarly,  $X_{20}$ and $X_{23}$ are signals of the second user during the $2^\text{nd}$ and $3^\text{rd}$ time slots.
\subsection{Coding Scheme for the Gaussian Channel}
The first user constructs its transmitted signals as
\noindent
\begin{align*}
X_{10}&=\sqrt{P_{10}}\check{X}_{10}(w_{10})+\sqrt{P_{U}}U(w_{12})\\
X_{13}&=\sqrt{P_{13}}\check{X}_{13}(w_{13})+\sqrt{c_2P_{U}}U(w_{12})+\sqrt{c_3P_{V}}V(w_{21})
\end{align*}
\noindent Similarly, the second user constructs its transmitted signals as
\noindent
\begin{align*}
X_{20}&=\sqrt{P_{20}}\check{X}_{20}(w_{20})+\sqrt{P_{V}}V(w_{21})\\
X_{23}&=\sqrt{P_{13}}\check{X}_{23}(w_{23})+\sqrt{d_2P_{V}}V(w_{21})+\sqrt{d_3P_{U}}U(w_{12})
\end{align*}
\noindent where $\check{X}_{10},\check{X}_{20},\check{X}_{13},\check{X}_{23},U,$ and $V$ are independent and identically distributed according to $N(0,1)$.

The power constraints for the two users are
\noindent
\begin{align*}
\alpha_1(P_{10}+P_{U})+\alpha_3(P_{13}+c_2P_{U}+c_3P_{V})&=P_1\nonumber\\
\alpha_2(P_{20}+P_{V})+\alpha_3(P_{23}+d_3P_{U}+d_2P_{V})&=P_2
\end{align*}
\noindent where $(c_2, c_3)$ are constant factors specifying the relative amount of power, compared to $P_{U}$ and $P_{V}$, used by the first user to transmit the cooperative information $(w_{12}, w_{21})$ during the $3^\text{rd}$ time slot. The same holds for $(d_2, d_3)$.
\subsection{Achievable Rate Region for the Gaussian Channel}
The achievable rate region for the half-duplex cooperative scheme over Gaussian channels can be expressed as in (\ref{Eq:rfin}) with the following expressions for $(I_2, I_4, I_5, I_6, I_{7})$:
\noindent
\begin{align*}
I_2&=\alpha_1 C\left(\frac{K_{12}^2\left(P_{U}+P_{10}\right)}{N_1}\right),\;I_4=\alpha_2 C\left(\frac{K_{21}^2\left(P_{V}+P_{20}\right)}{N_2}\right)\\
I_5&=\alpha_3 C\left(\frac{K_{10}^2P_{13}}{N_0}\right),\;\;\!\;\;\;\;\;\;\;\;\;\;\;\;I_6=\alpha_3 C\left(\frac{K_{20}^2P_{23}}{N_0}\right)\\
I_7&=\alpha_3 C\left(\frac{K_{10}^2P_{13}+K_{20}^2P_{23}}{N_0}\right)\;
\end{align*}
\begin{figure*}[t]
\normalsize
\begin{align}\label{Eq:bai}
I_8=&\;\alpha_1 C\left(\frac{K_{10}^2\mu_1}{N_0}\right)+
\alpha_3 C\left(\frac{K_{10}^2(P_{13}+c_2P_{U})+K_{20}^2(P_{23}+d_3P_{U})+2K_{10}K_{20}\sqrt{c_2d_3}P_{U}}{N_0}\right) \\
I_9=&\;\alpha_2 C\left(\frac{K_{20}^2\mu_2}{N_0}\right)+
\alpha_3 C\left(\frac{K_{10}^2(P_{13}+c_3P_{V})+K_{20}^2(P_{23}+d_2P_{V})+2K_{10}K_{20}\sqrt{d_2c_3}P_{V}}{N_0}\right) \nonumber \\
I_{10}=&\;\alpha_1 C\left(\frac{K_{10}^2\mu_1}{N_0}\right)+\alpha_2 C\left(\frac{K_{20}^2\mu_2}{N_0}\right)+\alpha_3 C\!\!\left(\frac{K_{10}^2P_{13}\!+K_{20}^2P_{23}\!+\!P_U(K_{10}\sqrt{c_2}+\!K_{20}\sqrt{d_3})^2\!+\!P_V(K_{10}\sqrt{c_3}+\!K_{20}\sqrt{d_2})^2}{N_0}\right)\nonumber
\end{align}
\vspace*{-8mm}
\end{figure*}
\noindent where $C(x)=0.5\text{log}(1+x)$ and $(I_8,I_9,I_{10})$ are given in (\ref{Eq:bai}) with $\mu_1=P_{10}+P_{U}$ and $\mu_2=P_{20}+P_{V}$.

\begin{figure}[t]
    \begin{center}
    \includegraphics[width=0.46\textwidth, height=62mm]{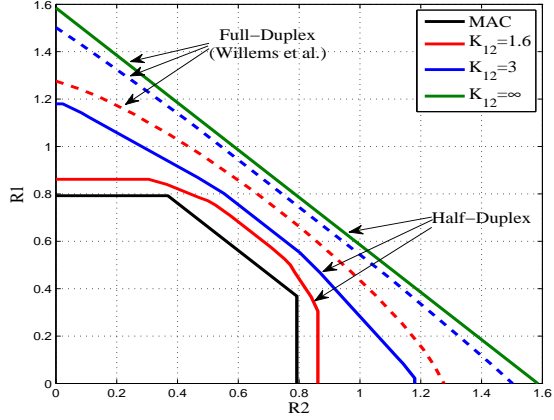}
    \caption{Achievable rate region for half-duplex cooperative scheme compared with full-duplex and classical MAC $(K_{10}=K_{20}=1, K_{12}=K_{21})$.} \label{fig:rate}
    \end{center}
\vspace*{-8mm}
\end{figure}

Fig. \ref{fig:rate} compares the achievable rate regions of the proposed half-duplex scheme, the full-duplex scheme in \cite{fcc2005fof,cognitiveRT}, and the MAC. The results are obtained for $N_0=N_1=N_2=1, P_1=P_2=2$, different values of $K_{12}$ and by using the optimal power allocations and time durations analyzed in \cite{haivu2}. Results show that our scheme has a larger rate region than the MAC, and the rate region increases as $K_{12}$ increases. As expected, our scheme has a smaller rate region than the full-duplex scheme. However, the two rate regions become closer to each other as $K_{12}$ increases.
\noindent
\section{Conclusion}\label{sec:conclusion}
In this paper, we have proposed a coding scheme for the half-duplex cooperative MAC based on superposition encoding
and ML decoding. The corresponding achievable rate region is derived. Numerical examples for the Gaussian channel
show that the achievable rate region improves with the inter-user channel quality. These results show that cooperation
can improve the rate region even with half-duplex constraint.
\appendices
\section{Proof of the Achievable Rate Region}

We want to find upper bounds for all error probabilities given in Section \ref{subsec:eranz}. We will follow the  procedure in \cite{haykin2005crb, abramowitz1972hmf}. First, we divide them into $4$ groups where error probabilities in each group have similar analysis as follows.
\noindent
\begin{itemize}
\item
$1^\text{st}$ group: $(P_{E_1}, P_{E_2}, P_{E_4}, P_{E_8})$
\item
$2^\text{nd}$ group: $(P_{E_3}, P_{E_5}, P_{E_6}, P_{E_7}, P_{E_9},$ $ P_{E_{10}}, P_{E_{11}})$
\item
$3^\text{rd}$ group: $(P_{E_{12}}, P_{E_{13}}, P_{E_{14}}, P_{E_{15}})$
\item
$4^\text{th}$ group: $(P_{E_{16}}, P_{E_{17}}, P_{E_{18}}, P_{E_{19}},P_{E_{20}})$
\end{itemize}
\noindent Since the analysis of the error events in the $4^\text{th}$ group are the most complicated, we provide here the full analysis of the $16^\text{th}$ error event. The analysis of the other error events in the same group or the other groups are not presented because of space constraint. However, they can evaluated similarly.
Now, we analyze the $16^\text{th}$ error event. Define $\vartheta_{w_{12}}$ as the event that (\ref{Eq:erst}) holds with  $\tilde{x}_1=x_1(1,1,1)$, $\tilde{x}_2=x_2(1,1,1)$, $\tilde{x}_{10}=x_{10}(1,1)$, $\tilde{x}_{20}=x_{20}(1,1)$, $\tilde{x}_{13}=x_{13}(1,1,1),$ and $\tilde{x}_{23}=x_{23}(1,1,1)$.
\begin{figure*}[t]
\normalsize
\begin{align}\label{Eq:erst}
P(y|x_1(w_{10},w_{12},w_{13}),x_2(1,1,w_{23}))\geq&\; P(y|\tilde{x}_1,\tilde{x}_2) \nonumber \\
\leftrightarrow P(y_1|x_{10}(w_{10},w_{12}))P(y_2|x_{20}(1,1))P(y_3|x_{13}(w_{13},w_{12},1),x_{23}(w_{23},w_{12},1))\geq&\; P(y_1|\tilde{x}_{10})P(y_2|\tilde{x}_{20})P(y_3|\tilde{x}_{13},\tilde{x}_{23}) \nonumber \\
\leftrightarrow P(y_1|x_{10}(w_{10},w_{12}))P(y_3|x_{13}(w_{13},w_{12},1),x_{23}(w_{23},w_{12},1))\geq&\; P(y_1|\tilde{x}_{10})P(y_3|\tilde{x}_{13},\tilde{x}_{23})
\end{align}
\vspace*{-8mm}
\end{figure*}
\noindent Then, the probability of this event is
\noindent
\begin{align*}
&P(\vartheta_{w_{12}})\!\!=\!\!\sum_{u}\sum_{x_{10}}\sum_{x_{13}}\sum_{x_{23}}\!\!P(u(w_{12}))P(x_{10}(w_{10},w_{12})|u(w_{12}))\times\\
&P(x_{13}(w_{13},\!w_{12},\!1)|u(\!w_{12}),\!v(1))P(x_{23}(w_{23},\!w_{12},\!1)|u(\!w_{12}),\!v(1))
\end{align*}
\noindent This probability can be bounded as \cite{haykin2005crb}
\noindent
\begin{align*}
P(\vartheta_{w_{12}})\!\!&\leq\!\!\sum_{x_{10}}\sum_{x_{13}}\sum_{x_{23}}\!\!P(x_{10}(w_{10},w_{12}))P(x_{13}(w_{13},w_{12},\!1)|v(1))\\
&\times P(x_{23}(w_{23},w_{12},\!1)|v(1))\left(\frac{P(y_1|x_{10}(w_{10},w_{12}))}{P(y_1|x_{10}(1,1))}\right)^s\\
&\times\left(\frac{P(y_3|x_{13}(w_{13},w_{12},1),x_{23}(w_{23},w_{12},1))}{P(y_3|x_{13}(1,1,1),x_{23}(1,1,1))}\right)^s.
\end{align*}
\noindent for any $s>0$. Now, let $\vartheta$ be the event that (\ref{Eq:erst}) holds for some $w_{12}\neq1$ and any $w_{10},w_{13},$ and $w_{23}$. Then for any
$0\leq\rho\leq1$, the probability of the event $\vartheta$ can be expressed as  shown in equation (\ref{Eq:akhr}) \cite{haykin2005crb, abramowitz1972hmf}.
For the ease of presentation, we use $(W_{12}, W_{10}, W_{13}, W_{23})$ to denote $(|W_{12}|, |W_{10}|, |W_{13}|, |W_{23}|)$.
\begin{figure*}[t]
\normalsize
\begin{align}\label{Eq:akhr}
\!\!\!\!P(\vartheta)\leq\!\!\left(\sum_{w_{12}=2}^{W_{12}}\sum_{w_{10}=1}^{W_{10}}\sum_{w_{13}=1}^{W_{13}}\sum_{w_{23}=1}^{W_{23}}\!\!\!P(\vartheta_{w_{12}})\!\!\!\right)^\rho  \!\!\!\!=\!W_E \!\left[\sum_{x_{10},x_{13},x_{23}}\!\!\!\!\!\!\!P(x_{10})P(x_{13}|v)P(x_{23}|v)
\left(\frac{P(y_1|x_{10})}{P(y_1|\tilde{x}_{10})}\right)^s\!\!\left(\frac{P(y_3|x_{13},x_{23})}{P(y_3|\tilde{x}_{13},\tilde{x}_{23})}\right)^s\right]^\rho
\end{align}
\vspace*{-11mm}
\end{figure*}
The probability of interest, $P_{E_{16}},$ has an upper bound as
\noindent
\begin{align*}
P_{E_{16}}\leq& \sum_{y_{13}}\sum_{x_{10}}\sum_{x_{13}}\sum_{x_{23}}\sum_{v}P(y_1|\tilde{x}_{10})P(y_3|\tilde{x}_{13},\tilde{x}_{23})\\
&P(\tilde{x}_{10})P(\tilde{x}_{13}|\tilde{v})P(\tilde{x}_{23}|\tilde{v})P(\tilde{v})P(\vartheta).
\end{align*}
\noindent By combining the last $2$ equations and choosing $s=1/(1+\rho)$ and $W_E=(W_{12}-1)^\rho W_{10}^\rho W_{13}^\rho W_{23}^\rho$, $P_{E_{16}}$ can be written as
\noindent
\begin{align*}
P_{E_{16}}\leq&W_E\sum_{y_{13}}\sum_{v}P(\tilde{v})\left[\sum_{x_{10}}P(\tilde{x}_{10})(P(y_1|\tilde{x}_{10}))^{\frac{1}{1+\rho}}\right]^{1+\rho}\\
&\times\left[\sum_{x_{13}}\sum_{x_{23}}P(\tilde{x}_{13},\tilde{x}_{23}|\tilde{v})(P(y_3|\tilde{x}_{13},\tilde{x}_{23}))^{\frac{1}{1+\rho}}\right]^{1+\rho}.
\end{align*}
\noindent Since the channel is memoryless, $P_{E_{16}}$ can be expanded as shown in (\ref{Eq:errd}) where $h=(\alpha_1+\alpha_2)n+1$. Then, by interchanging the order of the products and the summations, (\ref{Eq:errd}) becomes

\begin{figure*}[t]
\normalsize
\begin{align}\label{Eq:errd}
P_{E_{16}}\leq\; &W_E\!\!
\sum_{y_{1,1}^{\alpha_1n}}\sum_{y_{3,h}^{n}}\sum_{v_h^n}\prod_{i=h}^{n}\!\!P(v_i)\!\!  \left[\sum_{x_{10,1}^{\alpha_1n}}\prod_{i=1}^{\alpha_1 n}\!\!P(x_{10_i})(P(y_{1_i}|x_{10_i}))^{\!\frac{1}{1+\rho}}\!\!\right]^{\!\!1+\rho}\!\!\!\left[\sum_{x_{13,h}^{n}}\sum_{x_{23,h}^{n}}\prod_{i=h}^{n}
\!\!P(x_{13_i},x_{23_i}|v_i)(P(y_{3_i}|x_{13_i},x_{23_i}))^{\!\frac{1}{1+\rho}}\!\!\right]^{\!\!1+\rho}\\
\hline \nonumber
\end{align}
\vspace*{-12mm}
\end{figure*}
\noindent
\begin{align*}
&P_{E_{16}}\leq  W_E\prod_{i=1}^{\alpha_1 n}\sum_{y_{1_i}}\left[\sum_{x_{10_i}}P(x_{10_i})(P(y_{1_i}|x_{10_i}))^{\frac{1}{1+\rho}}\right]^{1+\rho}\times\\
&\prod_{i=h}^{n}\sum_{y_{3_i},v_i}\!\!\!P(v_i)\!\!\!\left[\!\!\sum_{x_{13_i},x_{23_i}}\!\!\!\!\!\!P(x_{13_i},x_{23_i}|v_i)(P(y_{3_i}|x_{13_i},x_{23_i}))^{\frac{1}{1+\rho}}\!\!\right]^{\!\!1+\rho}\!\!\!\!\!\!.
\end{align*}
\noindent Now, since the summations are taken over the inputs and the output alphabets, $P_{E_{16}}$ can be expressed as
\begin{align*}
&P_{E_{16}}\leq W_E L_1^{\alpha_1n}L_2^{\alpha_3n},\;\text{where}\\
&L_1\!=\!\!\sum_{y_{1_i}}\left[\sum_{x_{10_i}}P(x_{10_i})(P(y_{1_i}|x_{10_i}))^{\frac{1}{1+\rho}}\right]^{1+\rho}\\
&L_2\!=\!\!\!\!\sum_{y_{3_i},v_i}\!\!\!P(v_i)\!\!\!\!\left[\!\!\sum_{x_{13_i},x_{23_i}}\!\!\!\!\!P(x_{13_i},x_{23_i}|v_i)(P(y_{3_i}|x_{13_i},x_{23_i}))^{\frac{1}{1+\rho}}\!\!\right]^{\!\!1+\rho}\!\!\!\!\!.
\end{align*}
Following \cite{haykin2005crb}, $W_E$ has the following upper bound:
\noindent
\begin{align*}
W_E< 2^{n\left(R_{12}+R_{10}+R_{13}+R_{23}+\frac{2^{-n(R_{10}+R_{13}+R_{23})}}{(\text{ln}2)n}\right)}.
\end{align*}
\noindent Finally, the bound on $P_{E_{16}}$ can be expressed as follows.
\noindent
\begin{align*}
&P_{E_{16}}\leq 2^{-n\left[\Psi(\rho,P_{16})-\rho(R_{12}+R_{10}+R_{13}+R_{23})\right]},\;\text{where}\\
&\Psi(\rho,P_{16})=-\left(\alpha_1 \text{log}(q_1)+\alpha_3\text{log}(q_2)+\frac{2^{-n(R_{10}+R_{13}+R_{23})}\rho}{(\text{ln}2)n}\right)\\
&q_1\!=\sum_{y_{1_i}}\left[\sum_{x_{10_i}}P(x_{10_i})(P(y_{1_i}|x_{10_i}))^{\frac{1}{1+\rho}}\right]^{1+\rho}\\
&q_2\!=\!\!\sum_{y_{3_i},vi}\!\!\!P(v_i)\!\!\left[\!\sum_{x_{13_i},x_{23_i}}\!\!\!\!\!\!P(x_{13_i},x_{23_i}|v_i)(P(y_{3_i}|x_{13_i},x_{23_i}))^{\frac{1}{1+\rho}}\!\right]^{1+\rho}
\end{align*}
\noindent Now, it can be easily verified that $\Psi(\rho,P_{16})|_{\rho=0}=0$. Also,
\noindent
\begin{align*}
\left.\frac{d \Psi(\rho,P_{16})}{d\rho}\right|_{\rho=0}=&\;\alpha_1 I(X_{10};Y_1)-\frac{2^{-n(R_{10}+R_{13}+R_{23})}}{(\text{ln}2)n}\\
+&\;\alpha_3I(X_{13},X_{23};Y_3|V).
\end{align*}
\noindent Hence, it can be easily noted that $P_{E_{16}}\rightarrow 0$ as $n\rightarrow \infty$ if
\noindent
\begin{align*}
R_1+R_{23}<&\alpha_1 I(X_{10};Y_1)+\alpha_3I(X_{13},X_{23};Y_3|V).
\end{align*}
\noindent The rate constraints for the other error events are obtained similarly. Finally, after removing
the redundant constraints on the rates, we will have the achievable rate region given in (\ref{Eq:rnd}).
\bibliographystyle{IEEEtran}
\bibliography{references}
\end{document}